\documentclass[aps,prl, amsmath,amssymb,twocolumn,superscriptaddress,showpacs]{revtex4-1}

\usepackage{graphicx}
\usepackage{wrapfig}
\usepackage{appendix}
\usepackage{textcomp}
\usepackage{natbib}
\usepackage{amssymb}
\usepackage{subfig}
\usepackage{array}

\newcommand{\etal}{\emph{et al.}}
\begin{document}

\title{Improving photon detector efficiency using a high-fidelity optical CNOT gate}
\author{Katherine L. Brown}
      \affiliation{Department of Physics \& Astronomy and Hearne Institute of Theoretical Physics, Louisiana State University, Baton Rouge, LA, 70803}
	\author{Robinjeet Singh}
	\affiliation{Department of Physics \& Astronomy and Hearne Institute of Theoretical Physics, Louisiana State University, Baton Rouge, LA, 70803}
   \author{Joshua H. Mendez Plaskus}
   \affiliation{Department of Physics \& Astronomy and Hearne Institute of Theoretical Physics, Louisiana State University, Baton Rouge, LA, 70803} 
   \author{Hanna E. Broadus}
   \affiliation{Department of Physics \& Astronomy and Hearne Institute of Theoretical Physics, Louisiana State University, Baton Rouge, LA, 70803} 
\author{Jonathan P. Dowling}
	\email{jdowling@phys.lsu.edu}
   \affiliation{Department of Physics \& Astronomy and Hearne Institute of Theoretical Physics, Louisiana State University, Baton Rouge, LA, 70803} 
  
\date{\today}
   
\begin{abstract}    
A significant problem for optical quantum computing is inefficient, or inaccurate photo-detectors. It is possible to use CNOT gates to improve a detector by making a large cat state then measuring every qubit in that state. In this paper we develop a code that compares five different schemes for making multiple measurements, some of which are capable of detecting loss and some of which are not. We explore how each of these schemes performs in the presence of different errors, and derive a formula to find at what probability of qubit loss is it worth detecting loss, and at what probability does this just lead to further errors than the loss introduces. 
\end{abstract}

\maketitle
\section{Introduction}
Quantum computers exploit quantum superpositions to process every single input at the same time. However, when a measurement is performed, the quantum superposition collapses and only a single result is obtained. A standard quantum algorithm is run multiple times, before a meaningful data can be extracted. Despite this, there are several problems that can be solved significantly faster, using a quantum computer than using a classical computer. The standard example of this being Shor's factoring algorithm, which runs exponentially faster than any known classical factoring algorithm \cite{Shor1997}. Since each run of the quantum computer requires a significant amount of time, it is preferable to reduce the number of runs required. Depending on the context, there are two techniques for reducing the number of runs: the first technique exploits data extraction algorithms such as phase estimation \cite{Cleve1998}, and the second is the quantum state tomography \cite{Cramer2010}. While both of these techniques aim to minimize the number of runs of the algorithm required, they are both negatively impacted by inaccurate or missing measurements. Hence it is important to make the measurement procedure of our quantum computer as efficient and accurate as possible. 

Various achitectures proposed so far, for building a quantum computer, have their own advantages and disadvantages. A significant problem across these architectures is the efficiency and accuracy of the detector \cite{Hadfield2009, Atature2007, Morello2010,Schaetz2005}. In this paper we will explore how the efficiency of a detector can be improved using a miniature error correcting code. This miniature code can be layered on top of standard concatenated error corrections. We will concentrate on applying this miniature code to an optical quantum system, where our qubits are formed in the polarization basis of photons. In this scenario there is a unique problem that if a detector fails to make a measurement, the qubit is lost and it becomes impossible to detect its state.  Further we experience photon loss in the calculations, hence an ability to distinguish the photon loss occurring in the detector from the loss occuring in the calculations is important. To minimize the loss at the detector we assumed the detector efficiency of 90\%, which is optimistic compared to the currently available detectors \cite{Hadfield2009}.  

Previous work has explored a similar idea from a statistical perspective \cite{Deuar1999, Schaetz2005}. While Deuar and Munro \cite{Deuar1999} looked at a photonic system in a dual rail basis, Schaetz \etal~\cite{Schaetz2005} concentrated on an ionic system. We improve upon both the pieces of work in the following ways: First, we conduct full simulations rather than simple probabilistic calculations, therefore our results take into account any backwards propagation of error. Second, we propose several new adaptations to the standard scheme that are more suitable for architectures where there is loss. 

Depending on the fuction of the entangling gate, these adaptations are essential to distinguish the state $|0\rangle$ and the loss of the qubit that we are trying to measure. In their work, Ghosh \etal~\cite{Ghosh2013} discussed the topological errror correction advantages of using entangling gates for loss detection. 

In our present work, we propose five new schemes, as shown in table \ref{table1}, for improved detection. We developed a code that simulates these new schemes to be acting under following errors; probability that the measurement qubits are present, probability that the measurement qubits are initialized correctly, probability that the $X$ gate has an error, probability that the CNOT gate has an error, probability of loss in the $X$ gate, probability of loss and distribution of said loss in the CNOT gate, probability the detector looses the qubit, probability of a bit flip error in the detector, and the state and probability of loss of the initial photon. Our code can be used for any system but we chose error parameters most consistent with a photonic system. We then used our code to derive a formula to work out the loss probability when it becomes worth attempting to detect the loss, and the loss probability when loss leads to a greater error than the loss itself.


\newcolumntype{A}{>{\centering\arraybackslash} p{1.5cm} } 
\newcolumntype{C}{>{\centering\arraybackslash} m{8cm} } 
\newcolumntype{B}{>{\centering\arraybackslash} m{3.5cm} } 
\begin{table*}
\begin{tabular}{|A|C|B|}
\hline 
 & Description & Picture \\ 
\hline 
Scheme 1 & \begin{flushleft}A majority vote from the first $n$ readings.\end{flushleft} & \includegraphics[height=2cm]{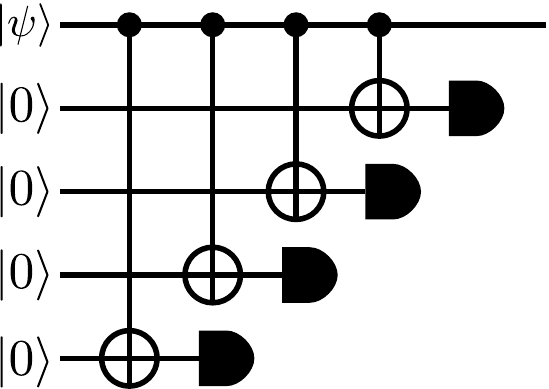} \\
\hline
Scheme 2 & \begin{flushleft}A majority vote from the first $n$ readings with an $X$ gate between each CNOT gate.\end{flushleft} & \includegraphics[height=2cm]{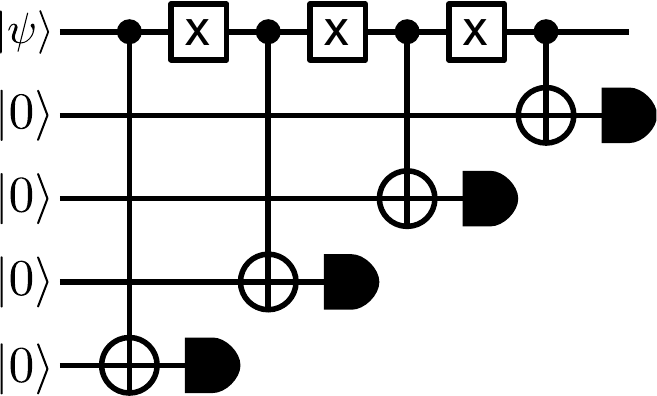} \\
\hline
Scheme 3 & \begin{flushleft}A majority vote where half the CNOT gates are performed then an $X$ gate is implemented and finally the other half of the CNOT gates are performed.\end{flushleft} & \includegraphics[height=2cm]{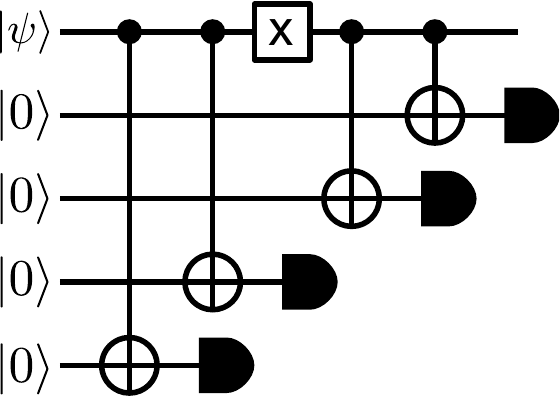} \\
\hline
Scheme 4 & \begin{flushleft}The first measurement to give a result is used as the reading.\end{flushleft} & \includegraphics[height=2cm]{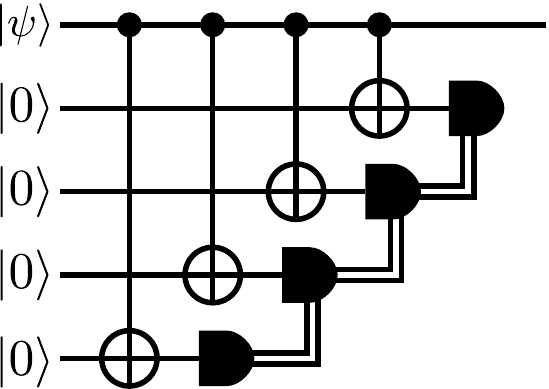} \\
\hline
Scheme 5 & \begin{flushleft}CNOT gates are used until one detector gives a result. At this point an $X$ gate is performed. Another chain of CNOT gates is then used until a second result is obtained. \end{flushleft}& \includegraphics[height=1.5cm]{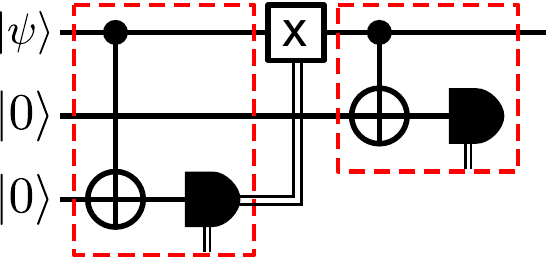} \\
\hline
\end{tabular}
\caption{We use five different measurement schemes, which are illustrated here for a total of four detectors.} 
\label{table1}
\end{table*}

In section II we first discuss the previous work  in comparison to our work, that has been done to compensate for inaccurate detection. We also introduce a statistical formula for the probability of obtaining the correct reading, in the presence of a CNOT gate with uncorrelated errors. In section III we introduce our error model that includes correlated errors on our CNOT gate, this something previous work has not taken into account. Then, in section IV we introduce our new five possible arrangements of CNOT gates, detectors and $X$ gates that can be used for our mini-error correcting code. In section V we discuss the simulation code we developed, and show the results it produces with the detector number as our variable. Section VI contains the results we obtained for deriving a formula to find the loss probability at which attempting to detect loss introduces less errors than the errors due to the loss itself. Finally, we conclude in section VII.

\section{Previous work} \label{previous}
Deuar and Munro \cite{Deuar1999} consider a copying device that is functionally equivalent to a CNOT gate in the vacuum--single-photon basis. Their aim is to determine photon presence in the presence of an inefficient detector and an error prone CNOT gate. If the detector has an efficiency given by $\eta$, and the copier has an error given by $\epsilon$ then as the number of measurements (N) tends to infinity, they find that the limiting efficiency is given by
\begin{equation}
\lim_{\text{N}\to \infty} \eta = 2-\frac{1}{\epsilon} \,.
\end{equation}
When $\eta=0.6$ and $\epsilon=0.714$ they find that three CNOT gates are needed before the cost of adding more copies no longer becomes worth the gain. Our model differs significantly from \cite{Deuar1999}. Our CNOT gate introduces errors onto the initial state being copied, which is physically more realistic and negatively impacts the maximum fidelity that can be achieved. It also implies that an infinite number of measurements would decrease the fidelity with respect to the optimal point. Further, Deuar and Munro \cite{Deuar1999} considered detection only in one basis, whereas our work consider distinguishing between three possible states, $|0\rangle$, $|1\rangle$ and loss. 

In their previous work on ionic qubit, Schaetz \etal~\cite{Schaetz2005} derived a formula for getting $m$ correct answers out of $M$ measurements  where the detection probability of the detector is given by $F$
\begin{equation}
P_{m}=\frac{F^m(1-F)^{M+1-m}(M+1)!}{m!(M+1-m)!} \,.
\end{equation}
To find the probability that the state is detected correctly they sum all the probabilities for $m>M/2$. This formula assumes a perfect CNOT gate, and that the error in the detector is a bit-flip rather than a loss error. An equivalent formula for an error-prone CNOT gate with a lossy detector (a scenario considered for photonic qubits) can be derived. The probability for the $m^{\text{th}}$ detector to give the correct answer is given by 
\begin{widetext}
\begin{equation}
C_{m}	= S.K\sum_{j=0}^{\lfloor m/2 \rfloor} \binom{m}{m-2j}([1-W][1-T])^ {2j}(W+T[1-W])^{m-2j}
\label{model}
\end{equation}
\end{widetext}
where $S$ is the probability that the second qubit is present, $K$ is the probability that the detector works. $[1-W][1-T]$ is the probability of detectable error in $|0\rangle-|1\rangle$ basis for $W$ to be the probability that the CNOT gate has no error, and $T$ to be the probability that any error in the CNOT gate is in the $Z$-basis, so that it has no impact on the measurement result in the $|0\rangle$, $|1\rangle$ basis. To find the probability of a correct majority vote we need to sum all scenarios where there are more correct readings than incorrect readings while ignoring the losses. 

The formula works on the principle that an even number of errors in the CNOT gates preceding the desired measurement lead to a correct answer, hence summing over $m/2$. For the assumption to work we need to consider errors on only one qubit, therefore assume that the CNOT error acts independently on each qubit that it occurs on. In reality this is not the case and we often have a correlated error model. This is one reason that we need to move from a statistical model to a simulation, and also that this model only works for a simple $|0\rangle$, $|1\rangle$ input and we want to consider the efficiency of a more general input. This general input becomes more important when we consider attempting to detect loss as the behavior of our CNOT gate, otherwise the loss will often give a false reading of $|0\rangle$. 

\section{Description of the error model} \label{error}
In this paper we use a standard Pauli model. The error model we use for the $X$ gate is given by 
\begin{equation}
\rho_{f}=\sum_{i}\sigma_{x}K_{i}\rho K_{i}^{\dagger}\sigma_{x}
\end{equation}
where $K_{0}=\sqrt{1-p_{x}}\,\mathbb{I}$ and $K_{n}=\sqrt{p_{x}/3} \sigma_{n}$ where $n=x,y,z$. 

The error model for the CNOT gate is given by
\begin{equation}
\rho_{f}=\sum_{i}\text{C}L_{i}\rho L_{i}^{\dagger}\text{C}
\label{CNOTe}
\end{equation}
where $C$ is the CNOT gate. Here we have $L_{0}=\sqrt{1-p_{c}}\,\mathbb{I}\otimes\mathbb{I}$ and $L_{i}=\sqrt{p_{c}/15}\,\sigma_{n}\otimes\sigma_{m}$ where $n=\mathbb{I},x,y,z$ and $m=x,y,z$ or $n=x,y,z$ and $m=\mathbb{I},x,y,z$. We note that equation (\ref{CNOTe}) results in correlated errors between the qubits that the CNOT gate acts upon. This means that the model we derived in the equation (\ref{model}) is not valid, we therefore look at full simulations rather than just the statistical model. If one of the qubits going into the CNOT gate is lost, we assume that the ideal CNOT gate performs an identity operation on the other qubit with the same error distribution as the standard CNOT gate. 

Here we use $|0\rangle$ to represent the horizontal polarization of a photon, and $|1\rangle$ to represent the vertical polarization of a photon. Our detection scheme assumes measurement in the $|0\rangle$--$|1\rangle$ basis. Since we are considering measurement at the end of a calculation, a necessary transformation can be applied for measurement to be made in the standard $Z$-basis.  We have not included the transformation error in the detection calculation, as we assume it to be the part of the computational error that is accounted for using standard error-correction or multiple runs of the computer. In quantum optical systems, the two-qubit gates require a nonlinearity that is typically introduced through measurement~\cite{Sipe}. However there are a few alternative ways of introducing this nonlinarity, such techniques include using a photonic module \cite{Devitt2007}, using the Zeno effect \cite{Leung2007} (although in this case measurement is still often required to improve efficiency)  and using a cross-Kerr nonlinearity as done in qubus systems \cite{Loock2008}.

\begin{figure}[h]
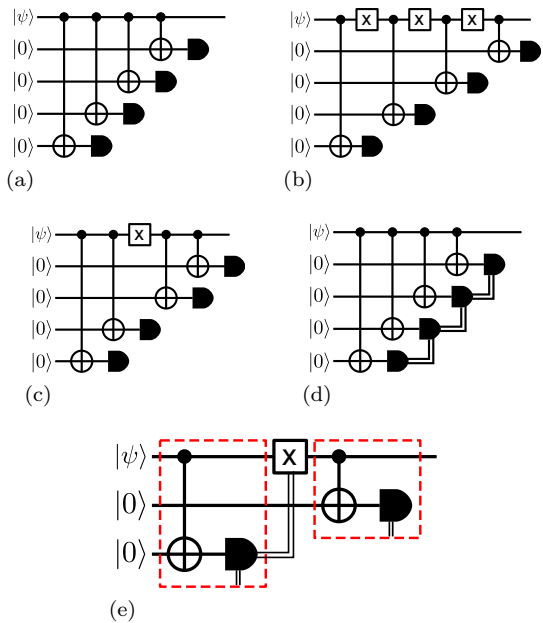

\captionsetup{justification=raggedright,singlelinecheck=false}
\subfloat[]{
\includegraphics[height=2cm]{One.pdf}
\label{one}}
\qquad
\subfloat[]{
\includegraphics[height=2cm]{Two.pdf}
\label{two}}
\qquad
\subfloat[]{
\includegraphics[height=2cm]{Three.pdf}
\label{three}}
\qquad
\subfloat[]{
\includegraphics[height=2cm]{Four.pdf}
\label{four}}
\qquad
\subfloat[]{
\includegraphics[height=2cm]{Five.pdf}
\label{five}}
\caption{Five possible schemes for detecting the state of a qubit in the $|0\rangle$, $|1\rangle$ basis. Circuits (a) and (d) have no loss detection, while circuits (b), (c) and (e) can detect loss.}
\label{alltechniques}
\end{figure}

\section{Alternative detection schemes} \label{alternative}

Schaetz \etal~\cite{Schaetz2005} discussed the fact that it should be possible to detect loss by using a simple bit flip operation (a Pauli $X$ gate) between the successive CNOT gates. In the case where the gates are error free then attempting to detect loss is clearly the best strategy. However, in reality both the CNOT and the $X$ gate will be subject to errors, and there is no reason to assume that these errors are identical. To detect the loss we need to use at least one additional $X$ gate and this results in a worse performance than a scheme which does not detect loss. As such we simulate five different techniques for improving our detector to see which one functions best under different error conditions. These schemes are illustrated in figure \ref{alltechniques} and summarised in table \ref{table1}. In this section we discuss each scheme and its advantages and disadvantages. 

Scheme 1, shown in figure \ref{one}, is the standard scheme of measurement with the CNOT gates. We make the assumption that the CNOT gates are performed on qubit one, sequentially, with no hold operations between them. The second qubit, in the CNOT gate, is prepared on demand and is measured straight after the CNOT gate. This technique has an advantage of using the minimal number of gates possible for a majority vote but has a disadvantage of being unable to detect when the initial qubit is lost. 

Schemes 2 and 3 use a combination of CNOT and $X$ gates. In scheme 2, as shown in figure \ref{two}, an $X$ gate is performed on qubit one after every CNOT gate. Scheme 3, as shown in figure \ref{three}, splits the total number of CNOT gates into two. First, half of the CNOT gates are performed that is followed by an $X$ gate on qubit one, and then the rest of the CNOT gates are performed. Both schemes 1 and 2 work best with an even number of CNOTs and detectors. Once again we perform the operations sequentially with no hold operations between them. The second qubit in each CNOT gate is prepared on demand and is measured straight after the CNOT gate. Both schemes 1 and 2 should be suitable for detecting loss, but if we assume a tight majority vote even a single error will mean that we will get a false reading of either $|0\rangle$ or $|1\rangle$.

Finally we consider schemes 4 and 5. Both of these schemes use the minimum number of CNOT gates and detectors possible. In scheme 4, as shown in figure \ref{four}, we perform CNOT gates and detections until one detector gives a reading. This reading alone is then used to determine our results. This scheme uses a considerably less number operations, on average, than required for a majority vote. Hence scheme 4 is more effective for CNOT gates with a high error rate. Similar to scheme 1, scheme 4 cannot detect the loss.  

Scheme 5, as shown in figure \ref{five}, is an adaptation of scheme 4 that is also able to detect the qubit loss. We consider performing CNOT gates and detections until a single measurement is made. Then, once this measurement has been made, an $X$ gate is performed on qubit one and we repeat our sequence of CNOT gates and detection. If we measure $|0\rangle|1\rangle$ then we conclude the first qubit was in $|0\rangle$, if we measure $|1\rangle|0\rangle$ then we conclude the first qubit was in $|1\rangle$. If the two measurements are the same, then we conclude the first qubit was lost. This technique has the advantage that it uses the minimum number of gates possible to detect loss.

\section{Number of detectors} \label{code}
To compare our five possible improved detector schemes, which are summarised in table \ref{table1}, we created a code that simulates the five schemes and compares the fidelity to the original input state. The fidelity calculation is computed over a three-level basis of $|0\rangle, |1\rangle$ and $|\text{loss}\rangle$. Our input state takes the form given by Eq \ref{eq}. Where $k_{1}$ is the probability that the qubit we are trying to detect is present, $k_{2}$ is the probability that the qubits used for detection are present, and $p_{0}$ is the probability that the photons used for detection are correctly initialized in $|0\rangle$. The wave-function $|\psi\rangle = \alpha|0\rangle + \beta|1\rangle$ is the state of the qubit we are trying to detect, and $|L\rangle$ is loss. Each measurement is modelled as a projective measurement, and is followed by a partial trace so that there is never any more than two qubits in the system. 

For each detection scheme we find the probability of each possible measurement combination. A simple summing procedure is then used to find the conclusion that an experimenter would draw from the set of measurement results. In the case of scheme 1, it simply involves using a dummy variable $S_t$ that starts at zero. Every time a $|0\rangle$ is measured one is added to $S_t$; every time a $|1\rangle$ is measured one is subtracted from $S_t$; when there is no detection $S_t$ is left constant. In the schemes with $X$ gates the procedure is flipped after every $X$ gate with a reading of $|0\rangle$ requiring the alternating addition or subtraction of one from $S_t$. We then sum the probability for all the combinations where $S_t$ is greater than zero, exactly zero, or less than zero and use this to form the probability of concluding $|0\rangle$, mixed, or $|1\rangle$, respectively. In the case of schemes 1 and 4, a mixed result is automatically an error; while in the cases of schemes 2, 3 and 5, a mixed result represents a conclusion of loss. 

\begin{figure*}[t]
\captionsetup{justification=raggedright,singlelinecheck=false}
\includegraphics[width=0.8\textwidth]{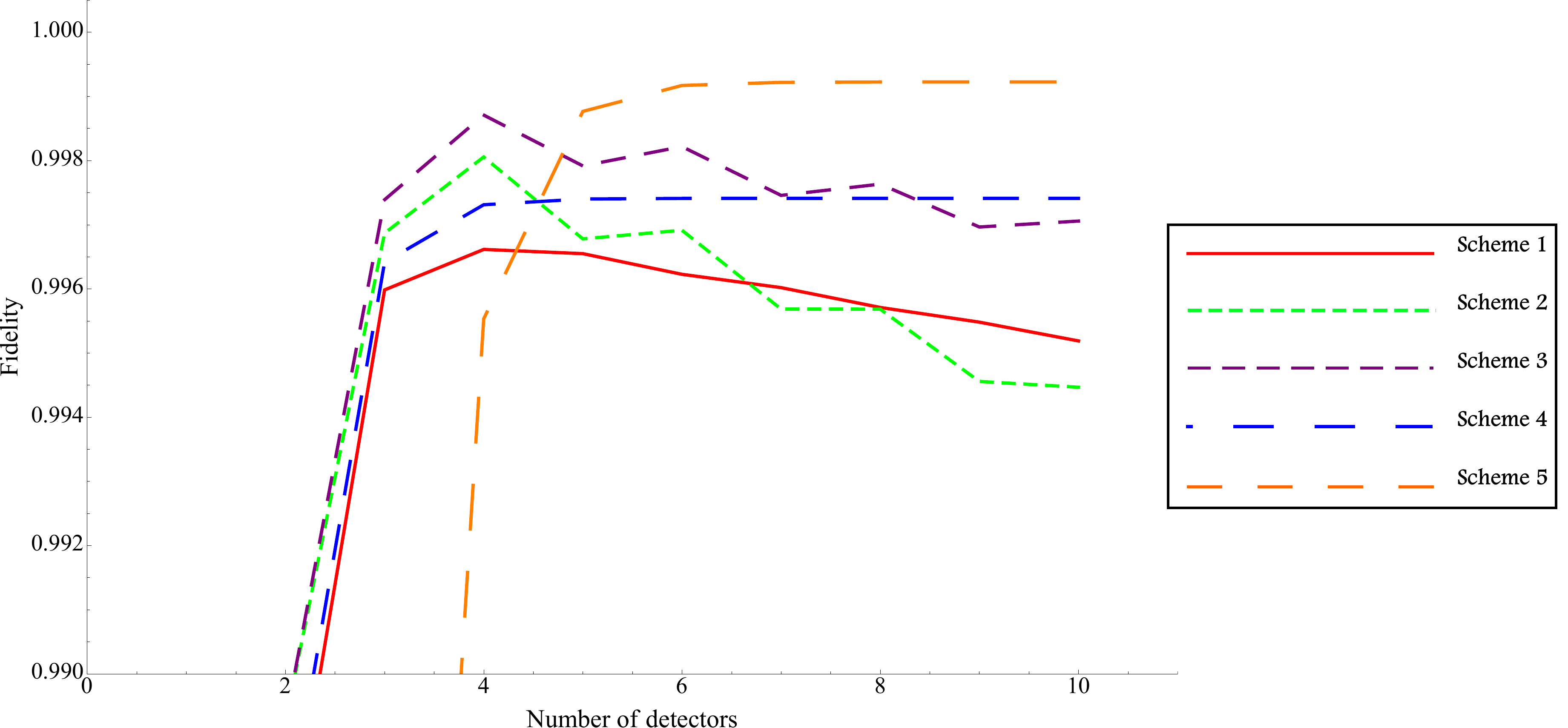}
\caption{How the five detection schemes discussed in section \ref{alternative} perform with respect to the number of detectors in the circuit. The qubit we are detecting is initially in the state $|1\rangle$ and always present, the detector error is $0.1$ while the Pauli error on the CNOT and $X$ gates is 0.001. The qubits used for detection are present with a $0.999$ probability, while the photon we are trying to detect is always present.}
\label{detectors}
\end{figure*}

\begin{widetext}
\begin{equation}
\psi_{\text{in}}=(k_{1}|\psi\rangle \langle\psi| + (1-k_{1})|L\rangle\langle L|) \otimes (k_{2}p_{0}|0\rangle \langle0| + k_{2}(1-p_{0})|1\rangle \langle1| + (1-k_{2})|L\rangle \langle L |)
\label{eq}
\end{equation}
\end{widetext}

If for some reason we want a system that has a higher chance of detecting the loss at an expense of a false positive for loss, it is possible to change the summation points. For example we could use $|S|\ge 1$ for the loss interval. We have not considered the higher loss detection scheme, in the present work, as our primary aim is to detect the state of our qubit in the $|0\rangle$, $|1\rangle$ basis. Therefore a false positive readings on loss would mean throwing away too much information. At the same time our code can easily consider these scenarios with a minimal adaptations. We note that in all cases, if our CNOT has a higher error in the $Z$-basis, than the detector, it is impossible to get any improvements.

Given that our CNOT error model means equation 3 is inaccurate, we want to look at how all five schemes perform as the number of detectors increases. An example case is shown in figure 2, where we start with our first qubit in the initial state $|1\rangle$ and have a Pauli error of $0.001$ on both the CNOT and $X$ gate, and a $0.1$ probability of loss in the detector. We chose a starting state of $|1\rangle$ because in schemes 1 and 4, where loss is not detected, any loss will always lead to a false reading of $|0\rangle$; we therefore consider the worst possible case. The photons used for detection are present with a $0.999$ probability. We see that scheme 4, where the minimum number of detectors is used, always outperforms the majority vote based scheme 1. This is because scheme 1 typically uses more CNOT gates. For a detector efficiency of $0.9$, we find that schemes 1$\textendash{4}$ reach a maximum fidelity at four detectors. Scheme 5, which uses the two first readings with an X-gate between them, has significant improvements for up to six detectors and after that increases only gradually. This is because the probability of obtaining at least two readings without loss increasing significantly until six detectors are used. Scheme 3, which only uses one $X$ gate, always outperforms scheme 2, which uses multiple $X$ gates; this is because the extra $X$ gates in scheme 2 provide no significant benefits but contribute to the net error.

A particular thing to note is the poor performance of scheme 5 relative to scheme 2 and 3 for a low number of detectors. This poor performance of scheme 5 is due to the fact that we have to have at least two detections before we get a reading, and a single CNOT or $X$-error will lead to a false reading of loss. Hence with four detectors we only expect a reading (including an incorrect one) with $0.999$ probability compared to the $0.9999$ probability of obtaining a reading for the other schemes. Similarly a single error on any of the gates will lead to an inaccurate reading in scheme 5, while the impact of an error on schemes 2 and 3 will depend on whether it occurred on the initial qubit (at which point it will effect all other measurements) or the qubit being detected. Therefore, despite the fact that taking the first reading (scheme 4) gives significant improvements over a majority vote (scheme 1), when there is no loss a majority vote (scheme 2) out-performs using the first two correct detections (scheme 5) for less than seven detectors. Unsurprisingly, when we increase the detector efficiency then scheme five begins to perform relatively well compared to schemes two and three, particularly in the cases of high loss. 

As we increase the detector efficiency, we reduced the optimal number of detectors needed. This is because each gate used introduces an error, so we want to use the minimum number of gates possible. Scheme 4 gets around this problem, to some degree, by using the minimum possible number of gates independent of the detector efficiency. Logically if we had no detector error then it would be better not to perform this enhanced measurement scheme and instead just measure the original qubit. Since we want to consider a range of detector efficiencies between $0.9$ and $1.0$, we will consider using four detectors for future calculations. This gives an accurate reading in the case of a 10\% detection without introducing too many errors in the case of low detector error. For higher detector errors more detectors would be required.

\section{When is it worth detecting loss?} \label{results}

\begin{figure*}[t]
\captionsetup{justification=raggedright,singlelinecheck=false}
\includegraphics[width=0.8\textwidth]{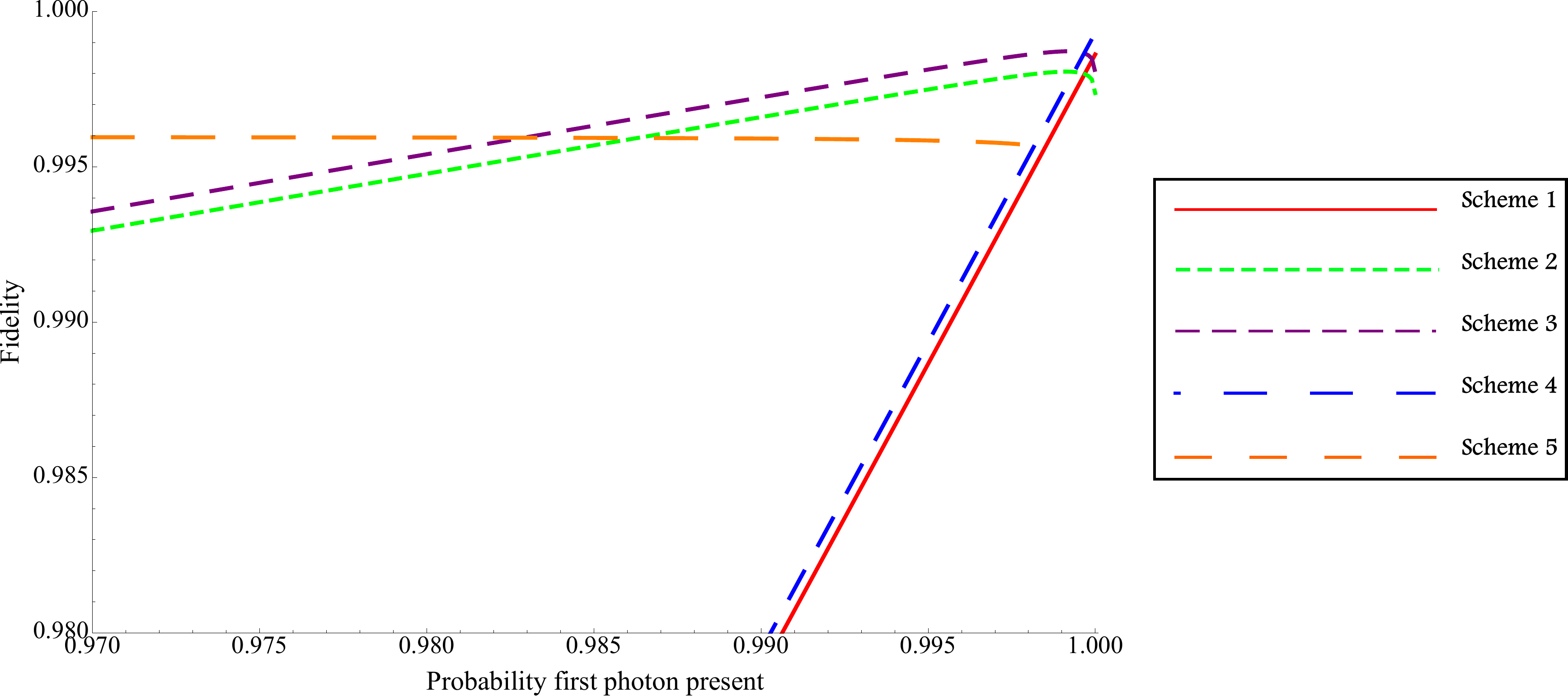}
\caption{Here we see how the five techniques perform in the presence of different values of loss. Here we have a detector error of $0.1$ and the probability of a Pauli error or loss error in the $X$ and CNOT gate is $0.001$. }
\label{Vloss}
\end{figure*}

In the previous section we decided that it was worth limiting the number of detectors we used to four. We now want to work out at what point is it worth detecting loss, and at what point does the increase in the number of gates required to detect the loss cause a greater error than the loss itself. From figure \ref{Vloss} we can see that using the first detector (scheme four) always outperforms a majority vote (scheme one) when we are using four detectors. We therefore consider this as our standard scheme for detecting the state of a qubit without loss. Unsurprisingly we see that scheme three, which only uses one $X$ gate, always outperforms scheme two which uses mutliple $X$ gates (the only exception is when we consider no error on the $X$ gate), in further results we therefore ignore scheme two. The relative performance of schemes three and five depend upon the level of loss as well as the gate error. In the cases of high loss, scheme five gives the highest fidelity, while in the cases of low loss, scheme three gives the highest fidelity. 

An interesting thing to note from figure \ref{Vloss} is that schemes two and three, which both use a fixed number of detectors and $X$ gates, show a dip in fidelity when the first photon has a $0.9991$ probability of being present. When the probability of having an error that can be read as loss is greater than the probability of correctly detecting loss, then the fidelity of our loss detection scheme will begin to fall. This happens when the ratio of the two errors, $R$, is greater than one. It is possible to approximate $R$ by
\begin{equation}
R=\frac{\zeta^{n-1} \times (1-\zeta)\times P_{p}}{\zeta^n \times (1-k_{1})} \Rightarrow \frac{(1-\zeta)\times k_{1}}{\zeta\times (1-k_{1})}
\end{equation}
where the probability that an $X$ or CNOT gate works is $\zeta=0.999$. To four significant figures, the probability where $R$ is greater than one is given by  $k_{1}=0.9991$. More generally we would expect the performance of schemes two and three to start decreasing when the probability of a gate error is higher than the probability of the photon being lost. If the error per gate is $0.0001$ then the performance of schemes two and three start dropping when $k_{1}=0.99991$, for an error of $0.0005$ the probability of presence where the schemes start to drop is $k_{1}=0.99949$,  for a $0.002$ error the drops happens at $k_{1}=0.9982$, while for an error of $0.005$ the performance starts to drop at $k_{1}=0.9957$. Therefore the formula above is very approximate; still it explains intuitively why we would expect the fidelity of the scheme to fall at roughly the point it does. 

We now derive a formula to show at what probability of loss, for the initial qubit, is it worth trying to detect loss. To do this we find the loss probability where scheme four, which takes the first reading without trying to detect loss, outperforms scheme three, a system that uses a fixed number of detectors and an $X$ gate in order to detect loss. The formula was derived by running our simulation code for values of CNOT error between $0.001-0.05$, $X$-error between $0.001-0.1$ and detector error between $0.001-0.1$, and using the Mathematica \emph{FindFit} function to find a formula that was linear in CNOT error and $X$-error and quadratic in detector error. The model was then checked with random values to ensure it was accurate to $\pm 0.0001$. The linearity in the $X$-error and CNOT error is not surprising since the two schemes differ by a constant number of CNOT and $X$ gates. The quadratic behavior in the detector errors is due to the fact that as the detector error increases the number of CNOT gates required by scheme four increases. Assuming that the gates have no loss error then the point where it becomes worth attempting to detect the loss is given by 

\begin{widetext}

\begin{equation}
\begin{aligned}
P_{L}=1 - [(0.0128122-0.575681P_{c} + P_{x}[-0.656495+3.75622P_{c}])P_{D}^2 +(-0.00117864+0.198512P_{c})P_{D} \\
					+ P_{x}P_{D}(0.115926-0.496572P_{c}) + 0.99986-0.203882P_{c} + P_{x}(-0.13992+0.41898P_{c})] 
						\label{result}
\end{aligned}						
\end{equation}
\end{widetext}

In a logical check we find that when all the errors are zero this gives $P_{L}=1-0.999986\pm0.0001$ which is within error bounds of $P_{L}=0$. The formula is only valid for $P_{D}\le 0.1$, $P_{x}\le 0.1$ and $P_{c}\le0.05$. This is not a strong restriction since with errors above these levels it is unlikely that quantum error correction will work, and therefore quantum computing would be impossible. For a more complicated error model our code can be used to find the best performing of the five error correcting schemes at any error probability and loss probability. 

\section{Conclusions} \label{conclusions}
In conclusion,  given a very good CNOT gate, detector and $X$-error we have derived formulas to find the optimal detecting scheme dependent upon the probability that the initial qubit is lost. We show that in a typical model with a CNOT and $X$-error error of $0.001$, and a detector error of $0.1$ it is worth attempting to detect loss if the probability that the qubit we are trying to detect is lost with a probability of greater than $0.0003$. If our CNOT and $X$ gate had errors of $0.05$ and the detector had an error of $0.1$ the probability of loss is increased to $0.0152$. A generalized formula is given in equation (\ref{result}).

To generate this model we have developed a code that compares all five schemes shown in table \ref{table1} given the following input errors; probability that the measurement qubits are present, probability that the measurement qubits are initialized correctly, probability that the $X$ gate has an error, probability that the CNOT gate has an error, probability of loss in the $X$ gate, probability of loss and distribution of said loss in the CNOT gate, probability the detector looses the qubit, probability of a bit flip error in the detector, and the state and probability of loss of the initial photon. This problem is difficult to solve analytically due to the correlated errors assumed in the CNOT gates, and the fact that loss during the process has different impacts depending on the initial state of our qubit. The code takes into account backward propagation of errors, and can be quickly be adapted to run over any variable and for any number of detectors. It is also possible to change how the majority vote is constructed, so that the loss detection can be made more or less cautious. Our model so far, is based on the projective error correction schemes and does not utilize other schemes like quantum feedback control, to take into account more complicated errors. Such calculations for further improved detector efficiency, are under ongoing investigation.

\begin{acknowledgements} 
Supported by the Intelligence Advanced Research Projects Activity (IARPA) via Department of Interior National Business Center contract number D12PC00527. The U.S. Government is authorized to reproduce and distribute reprints for Governmental purposes notwithstanding any copyright annotation thereon. Disclaimer: The views and conclusions contained herein are those of the authors and should not be interpreted as necessarily representing the official policies or endorsements, either expressed or implied, of IARPA, DoI/NBC, or the U.S. Government.
\end{acknowledgements}

\end{document}